\documentclass[conference]{IEEEtran}
\IEEEoverridecommandlockouts

\usepackage{url}
\usepackage{multirow}

\usepackage{cite}
\usepackage{amsmath,amssymb,amsfonts}
\usepackage{algorithm}%
\usepackage{algorithmicx}
\usepackage{algcompatible}
\usepackage{graphicx}
\usepackage{textcomp}
\usepackage{xcolor}
\usepackage{subcaption}
\usepackage{changes}
\usepackage{comment}
\def\BibTeX{{\rm B\kern-.05em{\sc i\kern-.025em b}\kern-.08em
    T\kern-.1667em\lower.7ex\hbox{E}\kern-.125emX}}

\newcommand{\bra}[1]{\langle #1|}
\newcommand{\ket}[1]{|#1\rangle}

\newcommand{\proj}[1]{\ket{#1}\!\bra{#1}}
\newcommand{\ketbra}[2]{\left|#1\rangle\langle#2\right|}

\newcommand{\abs}[1]{\left\lvert #1\right\rvert}

\newcommand{\beq}{\begin{equation}}
\newcommand{\eeq}{\end{equation}}

\def\cH{\mathcal H}

\newcommand\defn[1]{\textsl{#1}}

\newcommand{\bematrix}{\left(\begin{matrix}}
\newcommand{\ematrix}{\end{matrix}\right)}

\definechangesauthor[name={Giorgio Valentini}, color=blue]{GV}
\definechangesauthor[name={Mauricio Soto}, color=violet]{MS}
\definechangesauthor[name={Carlos}, color=red]{CC}

\begin{document}

\title{Hybrid Quantum-Classical Walks \\for Graph Representation Learning\\in Community Detection
}

\author{\IEEEauthorblockN{Anonymous Authors}}


\author{
\IEEEauthorblockN{1\textsuperscript{st} Adrián Marín}
\IEEEauthorblockA{\textit{QTCG.}\\
\textit{Dpt. Electromagnetism and Matter Physics.}\\
University of Granada, Spain \\
marinadrian@correo.ugr.es}
\and
\IEEEauthorblockN{2\textsuperscript{nd} Mauricio Soto-Gomez}
\IEEEauthorblockA{\textit{Dpt. di Informatica. } \\
Universit\`a degli Studi di Milano. Italy \\
mauricio.soto@unimi.it}
\and
\IEEEauthorblockN{3\textsuperscript{rd} Giorgio Valentini}
\IEEEauthorblockA{\textit{Dpt. di Informatica.} \\
Universit\`a degli Studi di Milano, Italy. \\
valentini@di.unimi.it}
\and
\IEEEauthorblockN{4\textsuperscript{th} Elena Casiraghi}
\IEEEauthorblockA{\textit{Dpt. di Informatica. } \\
Universit\`a degli Studi di Milano, Italy. \\
casiraghi@di.unimi.it}
\and
\IEEEauthorblockN{5\textsuperscript{th} Carlos Cano}
\IEEEauthorblockA{\textit{Dpt. of Computer Science and A.I.} \\
University of Granada, Spain \\
carloscano@ugr.es}
\and
\IEEEauthorblockN{6\textsuperscript{th} Daniel Manzano}
\IEEEauthorblockA{\textit{QTCG.}\\ 
\textit{Dpt. Electromagnetism and Matter Physics.}\\
University of Granada, Spain \\
dmanzano@ugr.es}
}

\maketitle

\begin{abstract}
Graph Representation Learning (GRL) has emerged as a cornerstone technique for analysing complex, networked data across diverse domains, including biological systems, social networks, and data analysis. Traditional GRL methods often struggle to capture intricate relationships within complex graphs, particularly those exhibiting non-trivial structural properties such as power-law distributions or hierarchical structures. This paper introduces a novel quantum-inspired algorithm for GRL, utilizing hybrid Quantum-Classical Walks to overcome these limitations. Our approach combines the benefits of both quantum and classical dynamics, allowing the walker to simultaneously explore both highly local and far-reaching connections within the graph. Preliminary results for a case study in network community detection shows that this hybrid dynamic enables the algorithm to adapt effectively to complex graph topologies, offering a robust and versatile solution for GRL tasks.

\end{abstract}

\section{Introduction}


The use of graph-based models to represent complex systems has become ubiquitous in diverse scientific domains, including biology, medicine, social sciences, and economics. This widespread adoption stems from the expressive power of graphs to capture intricate relationships among entities, such as interactions in protein-protein networks, patient similarities in medical records, or social connections in online platforms. As a result, Graph Representation Learning (GRL) — the task of learning compact, information-rich vector representations of graph nodes — has emerged as a central topic of interest in the machine learning community \cite{Hamilton2020}.

By embedding nodes into low-dimensional Euclidean spaces that preserve both local and global structural properties of the original graph, GRL enables the effective application of conventional machine learning models to tasks such as node classification, link prediction, community detection, and recommendation~\cite{Cui2019Survey, Goyal2018, zhang2018network}. 
Efforts in GRL leveraged a variety of algorithmic paradigms~\cite{khoshraftar2024survey}: ranging from matrix factorization techniques \cite{Lee1999} to graph neural networks (GNNs) \cite{Hamilton2020}, including Random Walk-based models such as DeepWalk and node2vec, which leverage local neighborhood sampling to learn representations based on co-occurrence statistics~\cite{Perozzi2014, Grover2016}.


In this work, we explore the use of Quantum Walks (QWs) to enhance random walk-based models for GRL. QWs have been an active line of research during the last decades \cite{kempe:cp03,aharonov:stoc01}. Most of its interest is based on the fact that QWs propagate quadratically, or even exponentially, faster than their classical counterparts for a huge variety of graphs \cite{kempe:ptrf05,makmal:pra14,makmal:pra16}. Besides, QWs are widely used to speed up computational problems such as searching \cite{shenvi:pra03,childs:pra14} and autonomous learning  \cite{paparo:prx14}. Interestingly, they have also played a foundational role in physics-inspired models of intelligence. For example, the Projective Simulation framework introduces a model of learning and decision-making based on stochastic and QWs on episodic memory graphs, offering a novel perspective on artificial agency rooted in quantum information processing~\cite{Briegel2012}. This connection highlights the versatility of walk-based mechanisms not only in representation learning but also in modeling cognitive-like processes in artificial agents and pinpoints a novel use for QWs in AI models. 

Beyond QWs, Hybrid Quantum-Classical Walks (HQCWs) mix both quantum and classical dynamics to determine the system's evolution. This kind of hybrid walks are grounded on the Open Quantum Systems (OQS) theory \cite{breuer_02},  and have been widely used to model different effects, such as photosynthetic complexes \cite{scholak:pre11,mohseni:jcp08,chin:njp10,manzano:po13}, computational science analysis \cite{sanchez-burillo:sr12}, and optimization of transport phenomena \cite{cao:jpc09,moix:njp13,Walschaers:prl13}. However, these approaches are based on quantum master equations that provide a description of the average behavior of a walker, lacking the information about single trajectories that is required for GRL techniques. 

In this paper, we provide an algorithm for the simulation of trajectories of HQCWs and a GRL technique based on it. This technique benefits from both the fast propagation of QWs and the deterministic trajectories obtained by the algorithm. We evaluate the potential of our model in the context of a community detection problem characterized by a complex graph composed of four random clusters of different sizes, with a low interconnectivity between them. This problem is difficult to handle with classical random walk-based methods since the different sizes of the clusters and the limited connections between them undermine their embedding representation capabilities. 

\section{Methods}

\subsection{Classical Walks}

Consider \(G = (V,E,w)\) to be a connected, weighted graph where  \(V = \{v_1,\ldots, v_N\}\) is the set of nodes of the graph, \(E = \{e_1,\ldots, e_M\}\) is the set of edges that connect the graph nodes, and \(w:{E}\to \mathbb{R}\)  is the weight function on the graph edges \cite{Newman}. A Classical Random Walk (CRW) on \(G\) is a discrete-time Markov chain \(\{X_t\}_{t \ge 0}\) taking values in the vertex set \({V}\). At each time step \(t\), if the walker is at node \(v_i \in {V}\), it moves to a neighboring node \(u \in N(v_i)\) chosen uniformly at random
\[
P(X_{t+1} = u \big| X_t = v_i ) = \tfrac{1}{k_{i}},
\]
where \(k_i = \deg(v_i)\).   Conveniently, we define an adjacency matrix \(A\) of \(G\) so that \(A_{ij} = 1\) iff \((v_i, v_j) \in {E}\) \cite{Norris_1997}. Consequently, \(k_i =  \sum_{j=1}^{V} A_{ij}\). This simple rule generates trajectories of any desired length without incorporating weight information or any bias or preference among nodes \cite{Perozzi2014}. 
To incorporate structural or homophilic biases, we can extend the CRW to a second‐order random walk in which the transition at time $t+1$ depends not only on the current node $v_i=X_t$, but also on the previously visited node $r=X_{t-1}$ \cite{Grover}. If $x \in N(v_i)$ and $w_{v_i,x}$ is the weight of the edge $(v_i,x)\in {E}$, we define an unnormalized transition score as 
\[
\pi_{rv_ix} \;=\;\alpha_{pq}(r,v_i,x)\,\cdot\,w_{v_ix},
\]
where the bias factor \(\alpha_{pq}\) is a function of the hop‐distance\footnote{The hop-distance $d_{ab}$ is defined as the minimum number of jumps needed to go from node $a$ to node $b$.} \(d_{rx}\) between the previous node \(r\) and the candidate \(x\):
\[
\alpha_{pq}(r,v_i,x)
\;=\;
\begin{cases}
\displaystyle\tfrac{1}{p} & \text{if } d_{rx}=0 \quad (\text{return to }r),\\[6pt]
1                        & \text{if } d_{rx}=1 \quad (\text{stay local}),\\[6pt]
\displaystyle\tfrac{1}{q} & \text{if } d_{rx}=2 \quad (\text{explore outward}).
\end{cases}
\]
Here \(p\) is the ``inward'' parameter controlling the likelihood of immediately returning to the previous node, and \(q\) is the ``outward'' parameter controlling the probability of exploring other parts of the graph \cite{sotogomez}. Finally, the normalized transition probability for this second‐order model is given by 
\[
P\bigl(X_{t+1}=x \mid X_t=v_i,\,X_{t-1}=r\bigr)
\;=\;
\frac{\pi_{rv_ix}}{\sum\limits_{z\in N(v_i)}\pi_{rv_iz}}\,.
\]
This flexible biasing scheme recovers the simple CRW when \(p=q=1\), but for other values of \((p,q)\) it can interpolate between highly local (BFS‐like) and far‐reaching (DFS‐like) walks on the same graph \cite{sotogomez}.

\subsection{Hybrid Quantum-Classical Walks}
Our proposed model is based on a Hybrid Quantum-Classical Walk. This means that the state of our system is given by a {\it quantum state}.  This corresponds to a column vector $\ket{\psi}$ (while $\bra{\psi}$ corresponds to a row vector) in a complex {\it Hilbert Space}\footnote{A Hilbert space is a vector space that is complete with respect to the metric induced by the inner product. For finite systems as the ones treated in this work a Hilbert space is just a vector space with an inner product.}, $\ket{\psi}\in \cH$. For our specific problem of a graph with $N$ nodes, the basis for our Hilbert space would be $\{ \ket{k}\}_{k=1}^N$, with $\ket{k}$ being the state in which the walker is at node $k$. Therefore, a general state of our HQCW can always be expressed as 

    \begin{equation}
        \ket{\psi} = \sum_{k=1}^{N}\, c_k\, \ket{k}\, ,
        \label{eq:vecexpansion}
    \end{equation}
where $c_k\in \mathbb{C}$ are the so-called amplitudes, and $\left| c_k \right|^2$ represents the probability of finding the particle at node $k$. From normalization, amplitudes should fulfill $\sum_k \abs{c_k}^2=1$. 

An alternative, yet equivalent, way of describing the state of a quantum system is the \defn{density matrix} formalism.   Consider such a system known to be in one of a set of $m$ possible quantum states $\{\ket{\psi_k}\}_{k=1}^m$ with corresponding probabilities $\{p_k\}_{k=1}^m$. The \defn{ensemble} 
$\{p_k, \, \ket{\psi_k}\}_{k=1}^m$ is described by the density operator $\rho = \sum_{k=1}^m p_k\, \ketbra{\psi_k}{\psi_k}$ \footnote{The expression $\ketbra{a}{b}$ represents a matrix obtained by the external product of the column  and the row vectors $\ket{a}$ and $\bra{b}$.  }. This kind of states, named {\it mixed states} combine the classical uncertainty of the imperfect determination of the state with the inherent quantum uncertainty due to the superposition principle.
%

Although interesting, purely QWs are not very useful for GRL. Due to the coherent character of the state, given by Eq. (\ref{eq:vecexpansion}), the walker is, in general, in several nodes at the same time. This makes it difficult to define a trajectory of the walker and to analyse its performance. To overcome this difficulty, we use a hybrid quantum-classical model that combines both coherent evolution with discrete jumps among nodes. This model has already been used in Ref. \cite{sanchez-burillo:sr12}  to analyse complex graphs. It was first proposed in Ref. \cite{schijven:jpa2012} to study the quantum to classical crossover in topologically disordered networks, and its transport properties were deeply studied in Ref. \cite{caruso:njp14}. However, as far as we know, our contribution is the first approach to describe how hybrid quantum-classical walks can be used for GRL. This hybrid model consists of a quantum system that evolves with both a coherent and an incoherent term. Its evolution is given by the Lindblad Master Equation that is the most general Markovian OQS master equation \cite{lindblad:cmp76,breuer_02,manzano:aip20}:
\begin{equation}
\frac{d\varrho}{dt} = -\mathrm{i}(1-\alpha)[H, \varrho] + \alpha\sum_{k,l}\left(L^{\phantom{\dagger}}_{k,l}\varrho L^\dagger_{kl} - \frac{1}{2}\left\{L^\dagger_{kl}  L^{\phantom{\dagger}}_{kl}, \varrho\right\}\right).
\label{eq:lindblad}
\end{equation}
In this equation, $i=\sqrt{-1}$, $[a,b]=ab-ba$ is the operators conmutator and $\left\{ a,b \right\}=ab+ba$ the anticonmutator. $H$ represents the Hamiltonian matrix that can be, in general, any Hermitian complex matrix. For our problem of QWs it is given by $H=\sum_{kl=1}^N
A_{kl} \left|k\rangle \langle l\right|$, 
where $k$ and $l$ are any two nodes in $V$ and $A$ is the adjacency matrix of the graph.
Similarly, the incoherent part is composed by jump operators in the form $L_{kl}= A_{kl} \left|k\rangle \langle l\right|$. The parameter $\alpha$ allows a continuous transition from a purely QW ($\alpha=0$) to a CRW ($\alpha=1$).

The Lindblad Master Equation gives the evolution of a classical ensemble, in a way similar to Markovian classical master equations describe the evolution of the probability distributions. To simulate single trajectories, we use a method called Quantum Montecarlo or Quantum-Jumps approach \cite{plenio:rmp98}. To simulate a walk with a depth of $L_{walk}$ jumps,  we apply the following algorithm: 
\begin{algorithm}[H]
\caption{Quantum-Jumps algorithm for HQCW}
\label{alg:vqa_adiabatic}
\begin{algorithmic}[1]
    \State Initial settings:\par
        \hskip\algorithmicindent $t=0$ (time) \par
        \hskip\algorithmicindent $\ket{\psi}=\ket{j}$ ($j$: initial node for the walk) \par
        \hskip\algorithmicindent $count=0$ (counter of jumps) 
    \State Determine the probabilities of jumping from  node $k$ to node $l$ for a time interval $\Delta t$ as \par
        \hskip\algorithmicindent $p_{kl}=\alpha \,A_{lk} \,\Delta t \left| \langle l |\psi \rangle \right|^2 \quad (k,\, l)\in [1,\,...,\,N]$
    \State Calculate the total probability of any jump as \par
        \hskip\algorithmicindent $P=\sum_{kl} p_{kl}$ 
    \State Uniformly sample $r\in[0,1]$ 
    \IF{$\mathbf{r<P}$}
            \hskip\algorithmicindent (there is a jump)
            \State Define the index $n\equiv k+N*(l-1)$. 
            \State Calculate the cumulative probabilities $P_n=\sum_{k=1}^n p_k$ 
            \State Find the value of $n$ such that $P_n\le r< P_{n+1}$ \par
        \hskip\algorithmicindent (this represents the jump made) 
            \State The state changes to $\ket{\psi}=\ket{l}$
            \State Store the site $l$ as visited 
            \State $count=count+1$
    \ELSE
        \hskip\algorithmicindent(there is no jump)
        \State The state evolves in a coherent (quantum) way as \par
        \hskip\algorithmicindent $\ket{\psi}\to\frac{\left\{ 1- (i/\hbar) H \Delta t - \sum_{kl} A_{lk} (\Delta t/2) \proj{k}  \right\} \ket{\psi} }{P}$. 
    
    \ENDIF
    
    \IF{$count<L_{walk}$}  
        \State Go to 2.
    \ENDIF

\end{algorithmic}
\end{algorithm}

In line 9 it is important to note that the value of $n$ unequivocally corresponds to a pair $(k,l)$; therefore, when we determine $n$ we are determining the pair of nodes where the jump is performed. In this model, when there is a jump (step 5 of the algorithm), we can consider that the walker has {\it collapsed} into this node and starts propagating again from it. This gives a Markovian character to the evolution, in a similar way to that for the CRWs. As the probabilities of the jumps are proportional to the parameter $\alpha$, it is clear that small values of this parameter make the system to propagate more coherently and with fewer jumps than higher values. At the limit $\alpha\to 0$ ($\alpha\to 1$), the quantum (classical) limit is recovered. Finally, the parameter $\Delta t$ is a numerical value that should be small enough to solve the Lindblad Master Equation (\ref{eq:lindblad}) numerically. The dependence of the results with regard to this parameter is not discussed in this work, and it will be subject to further research. 

\subsection{Node embeddings}

Both the HQCW and the Classical Methods (first- and second-order CRWs) can be applied to a graph $G$ to generate sequences of jumps/transitions that represent a serialization of G, ultimately yielding node embeddings.
Following the node2vec approach \cite{Grover}, we obtain node embeddings by learning continuous $d$-dimensional representations of the graph nodes through optimizing a skip-gram model \cite{mikolov2013distributed} by maximizing the concurrence probability among elements over the random‐walk sequences.

\section{Experimentation}

\subsection{Dataset and experimental settings}

\begin{figure}[h!]
    \centering    
    \includegraphics[width=0.8\linewidth]{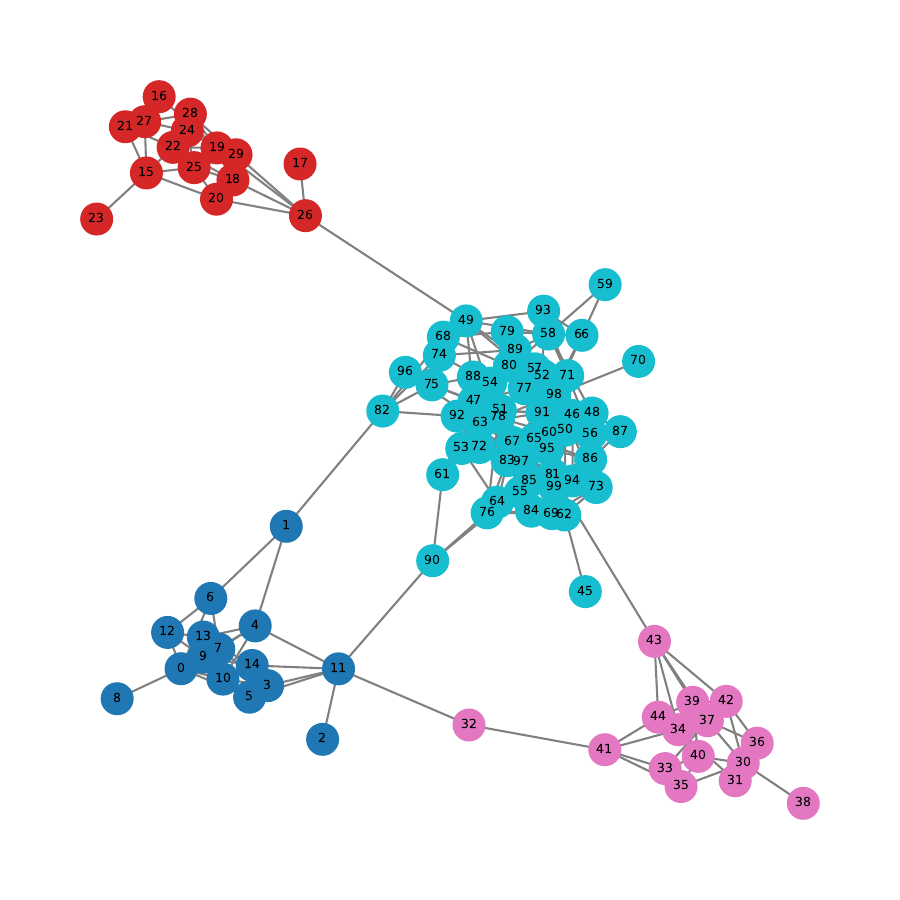}
    \caption{Graph analyzed in this work. Random graph of $100$ nodes with Erdös-Renyi subclusters of different sizes with  high intraconnectivity ($p_{\mathrm{intra}} = 0.25$) inside the clusters and small interconnectivity \(p_{\mathrm{inter}} = 0.0015\) between clusters. }
    \label{fig:graphs}
\end{figure}

Our studied graph is a random graph with dense subgraphs of different sizes as displayed in Fig \ref{fig:graphs}. It is composed of four Erdös-Renyi random networks of different scales: three of them are 15 nodes each, and one is 55 nodes. Within each subgraph, any given pair of nodes is connected with probability \(p_{\mathrm{intra}} = 0.25\). Between subgraphs, each possible edge is added with probability \(p_{\mathrm{inter}} = 0.0015\).

We evaluated the performance of the different methods in producing embedding representations capable of identifying the four clusters of different sizes. For each embedding, we applied K-means clustering with $K=4$ and retained the best solution among 50 consecutive runs, selecting the one with the lowest within-cluster sum of squares (inertia). For a quantitative assessment of the quality of the obtained clusters, we computed the Adjusted Rand Index (ARI). This is a statistical measure that quantifies the similarity between two clusterings of a dataset, correcting for the possibility of random agreement between partitions. The ARI is bounded below by -0.5 for especially discordant clusterings, it is ensured to have a value close to 0.0 for random labeling, and it is exactly 1.0 for identical clusters~\cite{hubert1985comparing}.  For a visual inspection of the node embeddings, we proposed the use of t-distributed stochastic neighbor embedding (t-SNE), a statistical method for visualizing high-dimensional data by projecting $\mathbb{R}^d$ vectors into two dimensions \cite{t-SNE}. For graphs with clear community structure, the 2D visualization of the resulting embeddings reveals how well nodes from the same cluster are grouped, enabling a qualitative assessment of the performance of each method.

The hyperparameters were set as follows. The classical walk-length \(L_{walk}\) was set to $10$, and for the Hybrid method this walk-length is equivalent to \(L_{walk} \simeq \alpha \; t_{total}\), where \(t_{total}\) is the total simulation time for each trajectory. The number of walks/trajectories to be performed starting from each node was set to 3. We tested embedding dimensions $d \in \{ 16,32, 64, 128\}$ and a context window size of 5. Second-order CRWs parameters were empirically set to $p=4$ and $q=0.1$.

The computational framework was implemented using Python libraries, including QuTiP for simulating quantum trajectories, NetworkX for graph handling, NumPy for numerical computations, Word2Vec for generating embeddings, Pandas for data processing and Scikit-learn for clustering the embeddings with K-means and computing the evaluation metrics. Both the datasets and the code used in this work are available at: \url{https://github.com/Adrianmarinb/HQCW-in-graphs}.

\subsection{Results and discussion}

\begin{figure}
    \centering
    \includegraphics[width=0.95\linewidth]{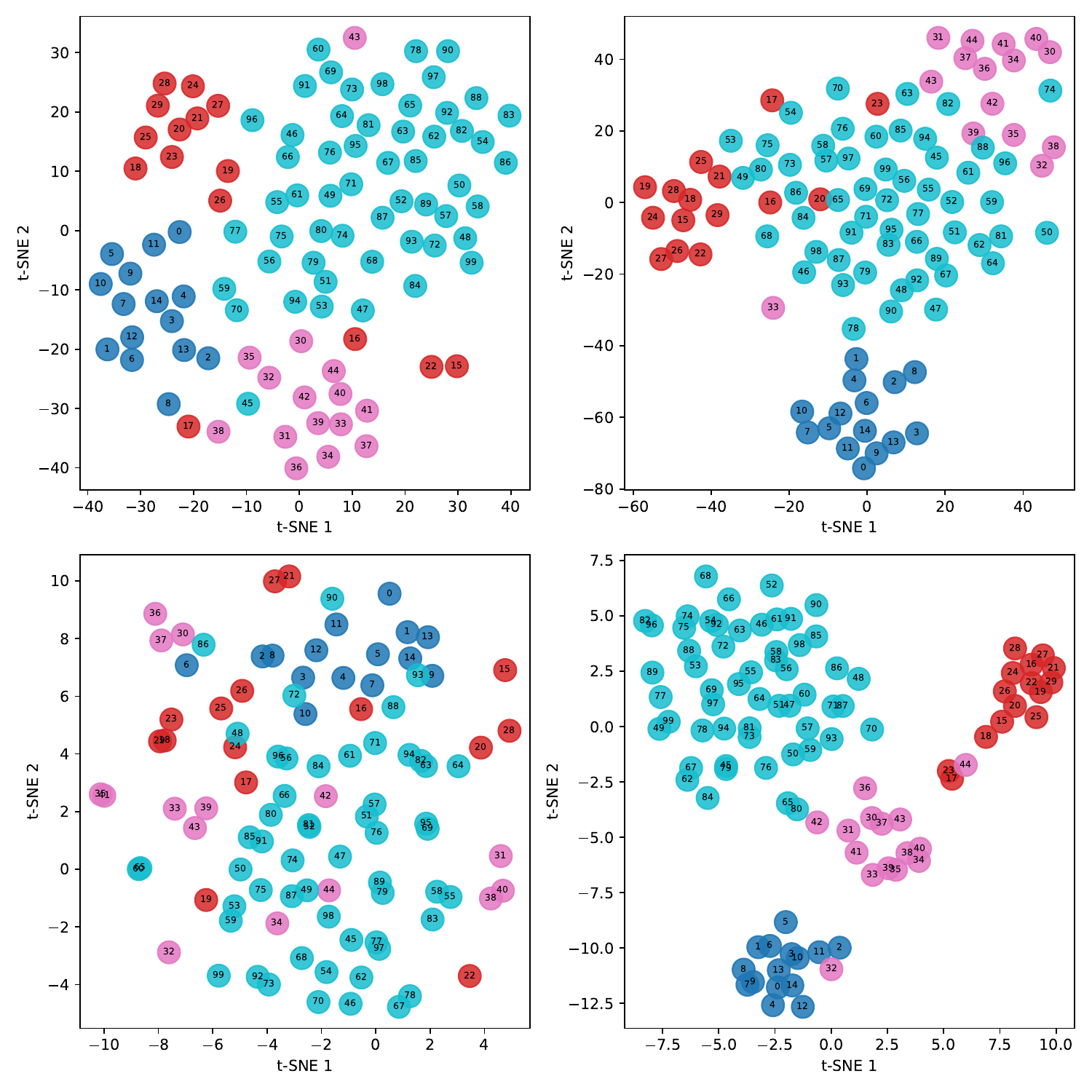}
    \caption{t-SNE representations of node embeddings ($d=32$) for CRWs (top): 1st order (left), 2nd order (right) and HQCWs (bottom): $\alpha=0.3$ (left), $\alpha=0.8$ (right).}
    \label{fig:clusters}
\end{figure}

Figure \ref{fig:clusters} depicts the t-SNE visualizations (with $d=32$) of node embeddings for both classical random walks (CRWs) and hybrid quantum-classical walks (HQCWs). These visualizations suggest that the first-order CRW and the highly quantum HQCW ($\alpha=0.3$) both fail to clearly separate the clusters, exhibiting only weak clustering with overlapping regions that hinder classification. In contrast, the second-order CRW improves classification compared to its first-order counterpart. Notably, the near-classical HQCW ($\alpha=0.8$) achieves even better separation, distinctly isolating clusters and enabling reliable classification. The ARI for these results is shown in Table~\ref{table1}. 
According to these results, a value of $\alpha=0.8$ is optimal for the proposed graph\footnote{Note that $\alpha=0.9$ presents a higher ARI but it uncertainty is also higher, making $\alpha=0.8$ the best choice of the parameter in a worst case scenario. }. This is consistent with previous findings~\cite{sanchez-burillo:sr12,schijven:jpa2012,caruso:njp14}. This value of $\alpha$ represents a walker that is almost classical, but still have some coherent transfer through the network. These results suggest a better performance of our model to classify complex clusters by spreading more effectively than first- and second-order CRWs. 

\begin{table}[]
\begin{center}
\begin{tabular}{c|c|c|}
\cline{2-3}
                                                                 & \(\alpha\)   & ARI                \\ \hline
\multicolumn{1}{|c|}{\multirow{7}{*}{HQCW}} & \(0.3\) & \(-0.022 \pm 0.010\) \\ \cline{2-3} 
\multicolumn{1}{|c|}{}                                     & \(0.4\) & \(-0.002 \pm 0.007\)   \\ \cline{2-3} 
\multicolumn{1}{|c|}{}                                     & \(0.5\) & \(-0.004 \pm 0.008\)   \\ \cline{2-3} 
\multicolumn{1}{|c|}{}                                     & \(0.6\) & \(-0.025 \pm 0.007\)   \\ \cline{2-3} 
\multicolumn{1}{|c|}{}                                     & \(0.7\) & \(-0.017 \pm 0.003\)   \\ \cline{2-3} 
\multicolumn{1}{|c|}{}                                     & \(0.8\) & \(0.030 \pm 0.002\)    \\ \cline{2-3} 
\multicolumn{1}{|c|}{}                                     & \(0.9\) & \(0.031 \pm 0.004\)    \\ \hline
\end{tabular}
\caption{ARI values of the node embeddings for the HQCWs for different values of the parameter $\alpha$.}
\label{table1}
\end{center}
\end{table}

\begin{table}[h!]
\begin{center}
\begin{tabular}{c|c|c|}
\cline{2-3}
                                                                 & \(d\)   & ARI                \\ \hline
\multicolumn{1}{|c|}{\multirow{4}{*}{2nd Order CRW}}             & \(16\)  & \(-0.004 \pm 0.002\) \\ \cline{2-3} 
\multicolumn{1}{|c|}{}                                           & \(32\)  & \(-0.009 \pm 0.008\) \\ \cline{2-3} 
\multicolumn{1}{|c|}{}                                           & \(64\)  & \(-0.004 \pm 0.000\) \\ \cline{2-3} 
\multicolumn{1}{|c|}{}                                           & \(128\) & \(-0.063 \pm 0.008\) \\ \hline
\multicolumn{1}{|c|}{\multirow{4}{*}{HQCW ( \(\alpha = 0.8\) )}} & \(16\)  & \(0.038 \pm 0.000\)  \\ \cline{2-3} 
\multicolumn{1}{|c|}{}                                           & \(32\)  & \(0.030 \pm 0.002\)  \\ \cline{2-3} 
\multicolumn{1}{|c|}{}                                           & \(64\)  & \(0.031 \pm 0.000\)  \\ \cline{2-3} 
\multicolumn{1}{|c|}{}                                           & \(128\) & \(0.022 \pm 0.000\)  \\ \hline
\end{tabular}
\caption{ARI values of the node embeddings for different dimensions of the embeddings \(d\) for the second-order CRW and HQCW $(\alpha=0.8)$.}
\label{table2}
\end{center}
\end{table}

\begin{figure*}
    \centering
    \includegraphics[width=0.95\linewidth]{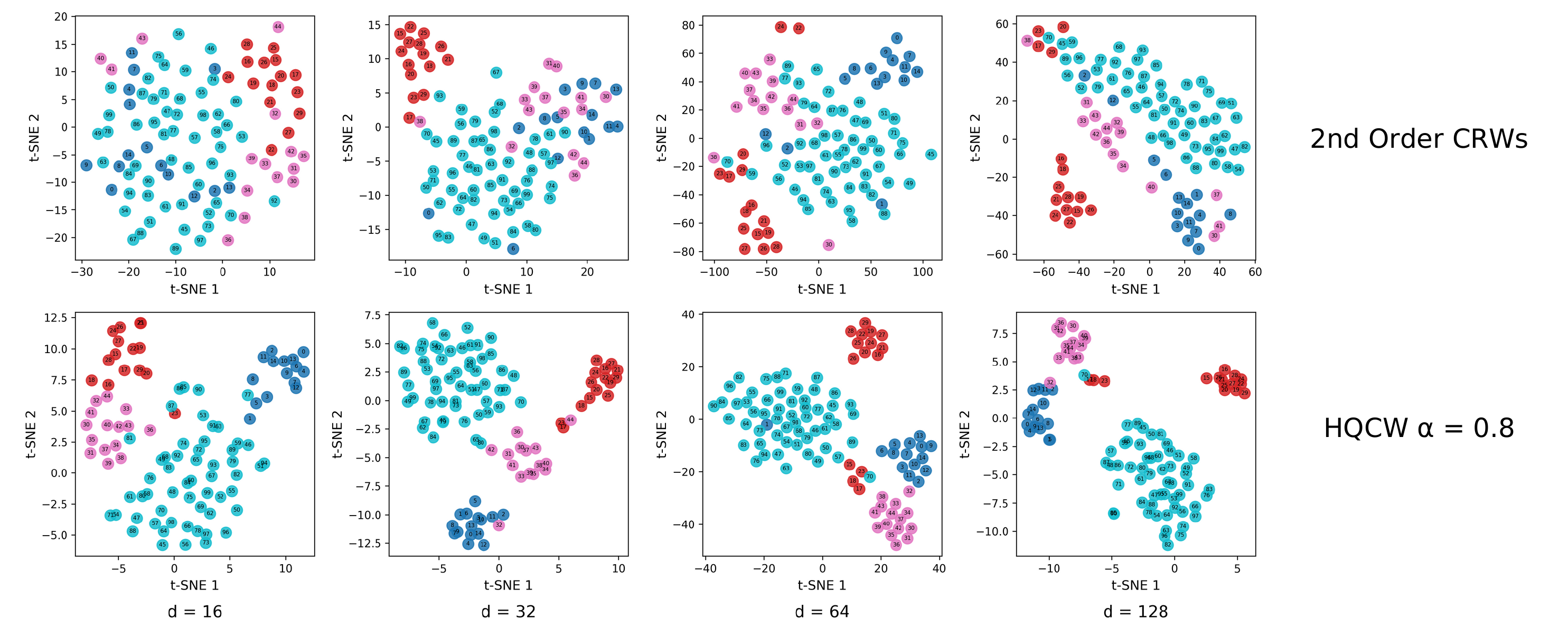}
    \caption{t-SNE representations for the 2nd order CRW (top row) and the HQCW with $\alpha=0.8$ (bottom row), for different embedding sizes $d$. }
    \label{fig:rows}
\end{figure*}

Further improvement in classification can be achieved by adjusting the embedding dimension $d$. Table~\ref{table2} shows the ARI values for the results obtained with the second-order CRW and the $\alpha=0.8$ HQCW across different values of $d$. These results show that the hybrid algorithm significantly outperforms the classical version for all values of $d$. In parallel, Figure \ref{fig:rows} compares the t-SNE representations of these embeddings. A visual inspection at these plots shows that, at a low dimension ($d=16$), the hybrid algorithm is significantly better than the classical version—a crucial result for scenarios where dimensionality constraints are imposed. As $d$ increases, the t-SNE visualization for the CRW embeddings improves, but it is always worse than the HQCW embeddings, for which the clusters become almost linearly separable with the t-SNE representation, enabling effective decision boundaries among them.

This analysis shows the superiority of our model for the proposed community detection problem. Although preliminary, these results show promising potential of HQCWs for node embedding and motivate further efforts and more extensive experimentation to fully assess their effectiveness for different applications.

\subsection{Complexity}
Finally, we have analised the computational complexity of the proposed HQCW model in comparison with CRWs: 

\begin{itemize}
    \item  {Space Complexity} 
    
    \begin{itemize}    
    \item CRWs: In its most basic form, implementing CRWs only requires storing the neighborhood of each node, which has a space complexity of $\mathcal{O}(|E|)$. For second-order random walks, transition probabilities can be efficiently calculated by storing the two-hop neighborhood of nodes, increasing the space requirement to $\mathcal{O}(\bar{d}^2|V|)$. $\bar{d}$ represents the average node degree, which is typically small in most practical applications \cite{Grover}.

    \item HQCWs: In general cases, implementing HQCWs requires storing both the Hamiltonian and Jump operators of Eq. \ref{eq:lindblad}. In our implementation, both coincide with the adjacency matrix $A_{kl}$. Storing this matrix has a complexity of $O(V^2)$. As this matrix is usually sparse, in the best case scenario it can be reduced to storing a set of $\mathcal{O}(|E|)$ as in the classical case.  
    \end{itemize}
    
    \item Time Complexity

    \begin{itemize}       
    
    \item CRWs: As a Markovian process, random walk generation allows for the reuse of samples. By generating a walk of length $L>l$, it becomes possible to derive $(L-l)$ random walks of length $l$ with a time complexity of $\mathcal{O}\left(\frac{L}{l(L-l)}\right)$ per sample \cite{Grover}.
    
    \item HQWRs: As HQWRs follow also a Markovian dynamics, the same argument holds giving an efficiency of 
$\mathcal{O}\left(\frac{L}{l(L-l)}\right)$ per sample.
    \end{itemize}

    \item Optimizations: The sampling process of both CRWs and HQWRs can be further enhanced through parallelization and the use of succinct data structures, as proposed in the literature for CRWs \cite{grape}. The use of these methods for the specific case of the HQWRs will be developed in further studies. 
\end{itemize}

\section{Conclusions}

In this paper, we have presented a novel model of quantum inspired random walks. This models uses the framework of Open Quantum System Theory in order to design a random walk that propagates faster than its classical counterparts. To test our model, we have analysed a graph composed of four sparsely connected Erdös-Renyi random networks of variable sizes. Community detection in this network is hindered by the large disparity between intra- and inter-community connectivity, which potentially limits the effectiveness of CRWs for this task.. We have found empirical evidence showing that the HQCWs can handle this problem in a better way than CRWs for all tested dimensions of the feature space of the embeddings. Consistent with prior studies, the best results for the HQCW method were obtained for $\alpha=0.8$. This represents a random walk that is almost classical, but with a small quantum transport. 

Some future directions of this problem that we plan to handle in the near future are to develop a ``discrete'' version of our algorithm in order to eliminate the $\Delta t$ parameter of our simulation, as well as to extend our analysis to further network topologies and problems, including power law networks and node ranking problems.

\section{Acknowledgements}

We acknowledge funding from MCUI/AEI (10.13039/501100011033) of the Spanish Government and FEDER, EU, through project PID2021-128970OA-I00 and QUANTUM ENIA project call - Quantum Spain project, and by the European Union through the Recovery, Transformation and Resilience Plan - NextGenerationEU within the framework of the Digital Spain 2026 Agenda. We also acknowledge project C-EXP-251-UGR23, cofunded by Consejería de Universidad, Investigación e Innovación and European Union call FEDER Andalucía 2021-2027. Finally, we acknowledge National Center for Gene Therapy and Drugs Based on RNA Technology—MUR (Project no. CN 00000041) funded by NextGeneration EU program.
We are also grateful for the the computing resources and related technical support provided by PROTEUS, the supercomputing center of Institute Carlos I for Theoretical and Computational Physics in Granada, Spain.

\bibliographystyle{IEEEtran}

\bibliography{biblio}

\end{document}